# DYNAMIC DUAL-GRAPH FUSION CONVOLUTIONAL NETWORK FOR ALZHEIMER'S DISEASE DIAGNOSIS

*Fanshi Li[1,2], Zhihui Wang[1,2], Yifan Guo[1,3], Congcong Liu[1,2], Yanjie Zhu[1,2], Yihang Zhou[1,3]*
*Jun Li[4], Dong Liang[1,3], Haifeng Wang[1,2]\**

[1] Paul C. Lauterbur Research Center for Biomedical Imaging, Shenzhen Institutes of Advanced Technology, Chinese Academy of Sciences, Shenzhen, Guangdong, China.
[2] University of Chinese Academy of Sciences, Beijing, China.
[3] Research Centre for Medical AI, Shenzhen Institutes of Advanced Technology, Chinese Academy of Sciences, Shenzhen, Guangdong, China.
[4] The Second People's Hospital of Shenzhen, Shenzhen, Guangdong, China.

## ABSTRACT

In this paper, a dynamic dual-graph fusion convolutional network is proposed to improve Alzheimer's disease (AD) diagnosis performance. The following are the paper's main contributions: (a) propose a novel dynamic GCN architecture, which is an end-to-end pipeline for diagnosis of the AD task; (b) the proposed architecture can dynamically adjust the graph structure for GCN to produce better diagnosis outcomes by learning the optimal underlying latent graph; (c) incorporate feature graph learning and dynamic graph learning, giving those useful features of subjects more weight while decreasing the weights of other noise features. Experiments indicate that our model provides flexibility and stability while achieving excellent classification results in AD diagnosis.

## 1. INTRODUCTION

Alzheimer's disease (AD) is the most widespread form of dementia, leading to memory loss or other cognitive issues due to damage to brain nerve cells [1]. Currently, AD affects approximately 6.2 million seniors aged 65 and over in the United States. This number is expected to increase to 13.8 million by 2060 [2]. Techniques like Magnetic Resonance Imaging (MRI) are efficient ways to track the advancement of Alzheimer's disease, and machine learning has been utilized to diagnose the disease from neuroimaging data [3].

Deep learning methods have made significant contributions to dementia diagnosis using MRI. Graph Convolutional Networks (GCNs) [4-5] in deep learning are an efficient way of incorporating diverse information and hold significant promise in the medical industry. Parisot et al. [6] used spectrum graph theory to perform the operations above on irregular graphs, improving simple KNN classifiers' performance by 11.9%. Kazi [7] has used a GNN network to learn category-specific multimodality features from the entire patient group and then used an attention mechanism to optimize integrating these multimodal features into the final decision-making. Subsequently, they [8] proposed a learnable function that can predict the optimal edge probability in the graph and achieve 94.14% accuracy in AD classification. Vivar et al. [9] utilized a combination of GCN with recurrent neural networks [10] to predict MCI-to-AD conversion with 87% accuracy. However, current GCN models still face some challenges in many medical problems. First, many current GCN models lack the ability to infer graph topology and are limited to a transductive setting, assuming that the graph structure is known and unchanging. This assumption is often unreliable as the graph may contain noise or be entirely unknown. Second, in medicine [11], samples are often challenging to gather, while the number of features per subject might quickly exceed thousands. Most normal GCN network struggles with this type of data, which can easily result in overfitting. Finally, some studies have started to make graph topology inferences [12], but the method's fundamental issue was that it modeled the graph in a completely linked manner without being able to take advantage of the graph's potential for sparsity and latent graph.

Motivated by the above, an innovative, dynamic dual-graph fusion convolutional network is proposed to detect AD using multi-modal MRI images. To this end, the proposed GCN architecture incorporates both interpretable feature graph learning and dynamic graph learning, intending to enhance diagnostic accuracy by learning the optimal latent graph structure through the adjustment of similar and dissimilar correlations across all subjects. As a result of co-optimizing feature graph learning, graph construction, and graph convolution, the proposed method for diagnosing diseases can achieve excellent diagnostic results comparable to other advanced classification methods.

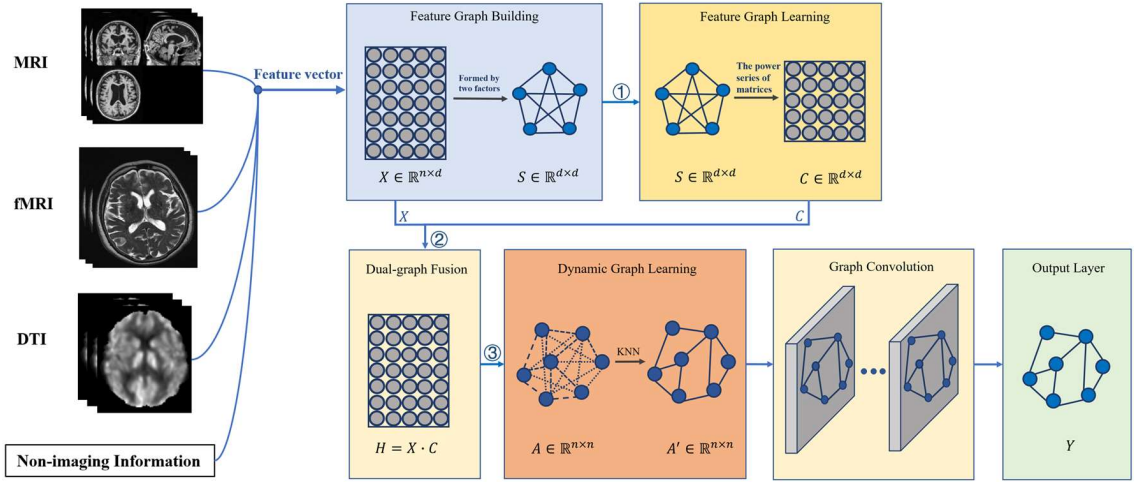

**Fig. 1.** Overview of the proposed model includes three modules, i.e., the module of Feature Graph Building, the module of Dynamic Graph Learning, and the GCN module.

## 2. PROPOSED AD DETECTION METHOD

### 2.1 Graph neural networks

An undirected graph $G = (V, E)$ can be used to represent this structure, where $V = \{v_1, v_2, ..., v_n\}$ is the vertex set and $E = \{e_1, e_2, ..., e_n\}$ is the edge set. The general graph representation is a six-tuple: $G(V, E, A, S, X, D)$, $A^{N \times N}$ is a (weighted) subject adjacency matrix, $S^{d \times d}$ is feature adjacency matrix, $X^{N \times d}$ represents the characteristic matrix of the node. The adjacency matrix $\tilde{A}$ is defined as $\tilde{A} = A + I$, where $I$ is the identity matrix. Diagonal degree matrix of $A^{N \times N}$ and $\tilde{A}^{N \times N}$ are denoted as $D^{N \times N}$ and $\tilde{D}^{N \times N}$, respectively. N and d represent the number of nodes and the feature dimension of nodes.

### 2.2 Model Architecture

In this section, we present our network for AD diagnosis, comprised of three modules: the feature graph learning module, the dynamic graph learning module, and the graph convolution module, as displayed in Figure 1.

*2.2.1. Feature Graph Learning:*
We built the module of feature graph learning. we use two factors to construct the feature graph $S$. The Fisher criterion is the first factor:

$$w_i = \frac{|u_{i,1} - u_{i,2}|^2}{\sigma_{i,1}^2 + \sigma_{i,2}^2} \quad (1)$$

where $u_{i,g}$ and $\sigma_{i,g}$ denote the mean and standard deviation, respectively, estimated for the $i$-th feature, based on the samples of the $g$-th class.

The second factor is the normalized mutual information $m_i$ between the class labels $Y$ and the features subjects of the $i$-th class $f_i$:

$$m_i = \sum_{y \in Y} \sum_{z \in f_i} p(z, y) \log \left( \frac{p(z, y)}{p(z)p(y)} \right) \quad (2)$$

where $p(\cdot, \cdot)$ symbolizes the joint probability distribution, $m_i$ represents the total reduction in uncertainty regarding a class that results from the information in the feature vector accumulated over all classes.

The two factors are combined through a linear weighting process:

$$s_i = w_i \alpha + m_i (1 - \alpha) \quad (3)$$

with $1 \leq i, j \leq n$. The coefficient $\alpha$ is a loading factor that falls within the range of [0,1], and experiments determine it. The weights in the adjacency matrix $S$ are obtained by associating the relevant $s$ through the following procedure:

$$\varphi_s(\vec{v_i}, \vec{v_j}) = S(i, j) = s_i s_j \quad (4)$$

From a mathematical perspective, Matrix $C$ can be calculated efficiently through the utilization of the convergence property of the geometric power series of a matrix [13]. Matrix $C$ represents the scores assigned to the features of subjects through encoding.

$$r = 0.9/\rho(S) \quad (5)$$
$$C = (I - rS)^{-1} - I \quad (6)$$

*2.2.2. Dynamic Graph Learning:*
We aim to assign higher weights to more relevant subject features while decreasing weights for features that do not contribute as much to the AD diagnosis. Therefore, we built the dynamic GCN by fusing $C$ and $X$ have $H \in \mathbb{R}^{N \times d}$ as:

$$H = XC \quad (7)$$

As the $C(i, j)$ can be seen as the length of the path enabled by feature $i$ to feature $j$ before the end of the selection process. It means that the more steps $i$ takes, the more associated it is with the feature nodes $j$. By marginalizing this quantity, we can get final relevance scores for each feature, the higher the score, the more important the feature $i$:

$$\tilde{c}(i) = [Ce]_i \quad (8)$$

Subsequently, we construct a graph adjacency matrix $A = \{a_{i,j}\}_{i,j=1}^{n}$ that reflects the correlation between subjects:

$$a_{i,j} = e^{-\theta \Delta(h_i, h_j)^2} \quad (9)$$

where $\theta$ present a learnable parameter and $\Delta(\cdot,\cdot)$ denotes the Euclidean distance. The edge weight $a_{ij} \geq 0$ denotes the affinity between subjects $h_i$ and $h_j$, which is multiplied by the feature attention matrix. Subsequently, the K-Nearest Neighbor (KNN) [14] method is utilized to generate a new unweighted adjacency matrix $A' \in \mathbb{R}^{N \times N}$.

Furthermore, we introduce a parameter $\lambda_1$ to facilitate the determination of the feature graph in the subsequent step. This parameter can be viewed as an implicit regularization mechanism that limits the deviation of the learned representation from the previous step (t) during a single operation.

$$H^t = X(C^{(t)} + \lambda_1 C^{(t-1)}) \quad (10)$$

*2.2.3. Graph Convolution Process:*

The node embeddings in GCNs are obtained through non-linear transformations and trainable weight matrices. In addition, GCNs utilize feature propagation to aggregate the data from a node's neighborhood. A conventional GCN layer is typically defined as follows:

$$H^{(l+1)} = \sigma(\widetilde{D}^{-\frac{1}{2}} A' \widetilde{D}^{-\frac{1}{2}} H^{(l)} W^{(l)}) \quad (11)$$

$H^{(l)}$ represents features from the $l$th layer, $W^{(l)}$ denotes the learnable weight matrix from the $l$th layer, and $\sigma$ is ReLU activation function.

The output matrix $H$ is obtained after the graph convolution process. The softmax function is applied to get the label prediction, represented as $Y = [y_0, y_1, \ldots, y_{n-1}] \in \mathbb{R}^{n \times c}$, where $y_i \in \mathbb{R}^c$ represents the prediction of the label for the $i$-th subject:

$$y_{ij} = softmax(h_{ij}^L) = \frac{\exp(h_{ij}^L)}{\sum_{m \in c} \exp(h_{im}^L)} \quad (12)$$

where $h_{ij}$ and $y_{ij}$ refer to the elements in the i-th row and j-th column of matrices, respectively.

*2.2.4. Loss function*

We utilized a composite loss function that incentivizes edges that contribute to accurate categorization and punishes those associated with incorrect categorization. The cross-entropy loss is calculated as:

$$L_{cross-entropy} = -\sum_{i \in N} y_i \ln z_i \quad (13)$$

where $y_i$ represents the ground truth of the $i$-th sample, while $z_i$ denotes the output of the $i$-th sample. Next we establish a reward function $\delta(z_i, y_i) = E((a_i)) - a_i$ that calculates the difference between the average accuracy of the $i$-th subject and the current success value $a_i = 1$ if $z_i = y_i$ and $a_i = 0$ if $z_i \neq y_i$. The graph loss is calculated as:

$$L_{graph} = \sum_{\substack{i=1 \ldots N \\ l=1 \ldots L \\ j:(i,j) \in \varepsilon^{(l)}}} \delta(z_i, y_i) \log a_{i,j}^{(l)} \quad (14)$$

Finally, the dynamic GCN model is optimized by minimizing the loss function below:

$$\mathcal{L} = \mathcal{L}_{cross-entropy} + \lambda_2 \mathcal{L}_{graph} \quad (15)$$

where $\lambda_2$ is the non-negative hyper-parameter used to balance the contribution of the two terms in the equation.

## 3. DATASET DESCRIPTION

The TADPOLE datasets, which are part of the Alzheimer's Disease Neuroimaging Initiative (ADNI) database [15], were utilized in our study and included data from 564 participants. Each participant's data is generated by integrating imaging features, such as fMRI, PET, and MRI images, with non-imaging information. The datasets consisted of 337 AD patients, 865 patients with Mild Cognitive Impairment (MCI), and 413 healthy controls (NC). Additionally, the MCI patient group was further divided into 315 MCI non-converters (MCIn) and 236 MCI converters (MCIp). The summarized information of all data sets can be found in Table 1. As the datasets have large feature dimensions, we used formula (8) to choose the top 60 features as the input.

TABLE I. Dataset distribution

| Data sets | Samples | Average age | Years in education | Female/Male | APOe4 |
|---|---|---|---|---|---|
| CN | 413 | 74.7 | 16.3 | 208/205 | 301/101/11 |
| MCI | 865 | 73.0 | 16.2 | 511/354 | 434/338/93 |
| AD | 337 | 75.0 | 15.2 | 186/151 | 114/158/65 |
| MCIn | 308 | 71.4 | 16.1 | 193/122 | 164/114/30 |
| MCIp | 216 | 73.4 | 16.4 | 131/85 | 76/119/21 |

## 4. EXPERIMENTS AND RESULT

### 4.1 Training Method

The performance of the proposed method was evaluated on our database using a 5-fold cross-validation strategy, and with the following parameters: a dropout rate of 0.1, regularization of $5 \times 10^{-4}$, a learning rate of 0.005, 50 epochs, a hyper-parameter k of 8 for the KNN method, and $\lambda_1$ values in the range of $[10^{-1}, 10^{-2}, \ldots, 10^{-6}]$ and $\lambda_2$ values in the range of $[0.2, 0.4, \ldots, 1]$. Evaluation metrics included prediction accuracy (ACC), sensitivity (SEN), specificity (SPE), and the area under the curve (AUC). The proposed method was tested in binary classification experiments, including AD vs. NC, AD vs. MCI, MCI vs. NC, and MCIn vs. MCIp, to validate our prediction performance.

### 4.2 Evaluation Method

In this study, the proposed prediction framework is compared with five widely used prediction frameworks: GCN [7], Random Forest (RF) [16], SVM [17], Adaboost [18], and DGM [8]. We performed individualized diagnoses on all datasets using different models and summarized the outcomes in Table 2 and Figure 2. In this section, we obtained the best classification results for the proposed model, with an ACC of 99.3% in the testing cohort. The results of our disease diagnosis method show an average improvement of 0.6% compared to the best comparison method (DGM) and an average improvement of 13.3% compared to the worst comparison method (SVM) in AD vs. NC task.

### 4.3 Feature selection analysis

In order to demonstrate the superiority of our proposed method, we compared its classification performance to two existing feature selection techniques, Recursive Feature Elimination (RFE) [19] and ANalysis Of VAriance (ANOVA) [20]. The comparison results, depicted in Fig. 3, demonstrate that our method surpasses the comparison methods in classification accuracy. Specifically, our method achieved a roughly 2.3% improvement in accuracy compared to RFE on the MCIn-MCIp dataset. Furthermore, we also obtained the most important features selected using our method in AD vs. NC according to Eq. (8). It is observable that the features are closely correlated to AD [21]. For instance, on the clinical data, there are Entorhinal, Parahippocampal, Hippocampus, Supramarginal, etc., and on the non-clinical data, these are CDRSB, APOE4, and RAVLT_learning, etc.

### 4.4 Parameters' sensitivity analysis

Figure 4 illustrates the fluctuation of the classification accuracy because of adjusting the values of the two hyper-parameters in Eq. (10) and Eq. (16), i.e., $\lambda_1$, and $\lambda_2$. Our proposed model achieves the best performance with $\lambda_1 = 10^{-2}$ and $\lambda_2 = 1$ on AD-NC and AD-MCI, $\lambda_1 = 10^{-4}$ and $\lambda_2 = 1$ on NC-MCI, $\lambda_1 = 10^{-3}$ and $\lambda_2 = 0.8$ on MCIn-MCIp. It is evident that the proposed method is susceptible to the choice of parameters, as evidenced by the fluctuation of 6.7% in classification accuracy on some datasets. Compared to the parameters $\lambda_1$, $\lambda_2$ has a greater influence on the final classification outcome as it is responsible for regulating the performance of the final classification process.

## 5. CONCLUSION

In this paper, we applied a dynamic dual-graph convolutional neural network to address the issue of GCN. Our framework proposes a dynamic graph learning approach for AD diagnosis that is suitable for multi-modal datasets and can be independently evaluated on datasets. Experiment results on the ADNI datasets validate the excellent performance of our proposed model when compared to the other advanced methods. Later, we intend to broaden our method for interpretable feature extraction and unbalanced datasets with few samples.

Table II. Classification results of all methods

| Dataset | Metrics | SVM | Ada-boost | RF | GCN | DGM | Proposed |
|---|---|---|---|---|---|---|---|
| AD vs NC | ACC | 86.0 | 90.0 | 90.9 | 92.1 | 98.7 | **99.3** |
| | SEN | 60.0 | 80.0 | 60.0 | 90.0 | 97.0 | **98.7** |
| | SPE | 93.3 | 93.3 | 100 | 96.3 | 100 | **100** |
| | AUC | 99.4 | 86.3 | 99.4 | 95.8 | 99.1 | **99.8** |
| AD vs MCI | ACC | 76.6 | 90.0 | 80.0 | 88.5 | 92.5 | **94.6** |
| | SEN | 50.0 | 83.3 | 76.6 | 90.1 | 97.1 | **96.5** |
| | SPE | 100 | 93.3 | 81.6 | 81.6 | 80.5 | **89.5** |
| | AUC | 90.8 | 84.8 | 90.1 | 89.4 | 95.3 | **98.0** |
| MCI vs NC | ACC | 83.9 | 86.7 | 83.3 | 90.5 | 95.3 | **98.0** |
| | SEN | 89.0 | 70.0 | 73.3 | 89.2 | 95.3 | **100** |
| | SPE | 73.2 | 0.93 | 86.7 | 91.4 | 95.1 | **93.9** |
| | AUC | 91.7 | 82.8 | 89.7 | 91.2 | 98.0 | **99.1** |
| MCIn vs MCIp | ACC | 76.0 | 72.0 | 84.0 | 74.5 | 81.0 | **88.3** |
| | SEN | 76.6 | 73.3 | 83.3 | 74.8 | 68.7 | **79.2** |
| | SPE | 73.3 | 66.7 | 83.3 | 83.6 | 89.4 | **94.7** |
| | AUC | 83.3 | 81.3 | 89.5 | 81.4 | 87.5 | **94.2** |

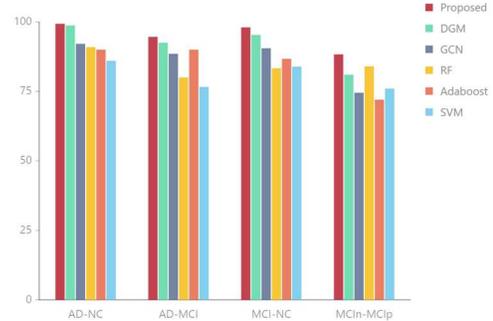

**Fig. 2. Classification results of all methods on four datasets.**

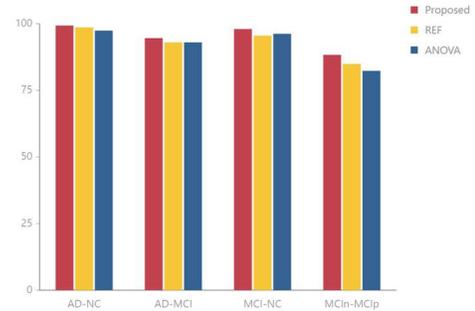

**Fig. 3. The classification accuracy of three feature selection methods.**

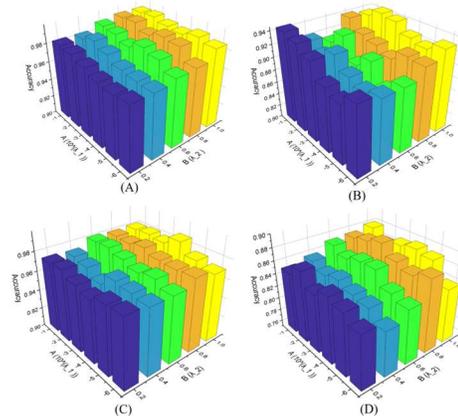

**Fig. 4. Our method with different parameter settings (i.e., $\lambda_1$ and $\lambda_2$) on (A) AD-NC, (B) AD-MCI, (C) NC-MCI, (D) MCIn-MCIp.**

# 6. REFERENCES

[1] S. Karantzoulis and J. E. Galvin, "Distinguishing Alzheimer's disease from other major forms of dementia," *Expert Review of Neurotherapeutics*, vol. 11, no. 11, pp. 1579–1591, Nov. 2011.

[2] "2021 Alzheimer's disease facts and figures," *Alzheimer's & Dementia*, vol. 17, no. 3, pp. 327–406, Mar. 2021.

[3] X. Zhu et al., "Joint prediction and time estimation of COVID-19 developing severe symptoms using chest CT scan," *Medical Image Analysis*, vol. 67, p. 101824, Jan. 2021.

[4] M. Gori, G. Monfardini, and F. Scarselli, "A new model for learning in graph domains," in *Proc. Int. Joint Conf. Neural Netw.*, 2005, pp. 729–734.

[5] F. Scarselli, M. Gori, A. C. Tsoi, M. Hagenbuchner, and G. Monfardini, "The graph neural network model," I*EEE Trans. Neural Netw.*, vol. 20, no. 1, pp. 61–80, Jan. 2009.

[6] S. Parisot et al., "Spectral Graph Convolutions for Population-Based Disease Prediction," in *Medical Image Computing and Computer Assisted Intervention − MICCAI 2017*, vol. 10435, 2017, pp. 177–185.

[7] A. Kazi et al., "Graph Convolution Based Attention Model for Personalized Disease Prediction," in *Medical Image Computing and Computer Assisted Intervention − MICCAI 2019*, vol. 11767, 2019, pp. 122–130.

[8] A. Kazi, L. Cosmo, S.-A. Ahmadi, N. Navab, and M. Bronstein, "Differentiable Graph Module (DGM) for Graph Convolutional Networks," *IEEE Trans. Pattern Anal. Mach. Intell.*, vol. 45, no. 2, pp. 1606–1617, Feb. 2023.

[9] G. Vivar, A. Zwergal, N. Navab, and S. A. Ahmadi, "Multi-modal disease classification in incomplete datasets using geometric matrix completion," in *Proc. Med. Image Comput. Comput. -Assist. Intervention*, 2018, pp. 24–31.

[10] W. Zaremba, I. Sutskever, and O. Vinyals, "Recurrent Neural Network Regularization." *arXiv*, Feb. 19, 2015.

[11] D. Dernoncourt, B. Hanczar, and J.-D. Zucker, "Analysis of feature selection stability on high dimension and small sample data," *Computational Statistics & Data Analysis*, vol. 71, pp. 681–693, Mar. 2014.

[12] L. Cosmo, A. Kazi, S.-A. Ahmadi, N. Navab, and M. Bronstein, "Latent-Graph Learning for Disease Prediction," in *Medical Image Computing and Computer Assisted Intervention − MICCAI 2020*, vol. 12262, 2020, pp. 643–653.

[13] G. Roffo, S. Melzi, U. Castellani, A. Vinciarelli, and M. Cristani, "Infinite Feature Selection: A Graph-based Feature Filtering Approach," *IEEE Trans. Pattern Anal. Mach. Intell.*, vol. 43, no. 12, pp. 4396–4410, Dec. 2021.

[14] T. Cover and P. Hart, "Nearest neighbor pattern classification," *IEEE Trans. Inform. Theory,* vol. 13, no. 1, pp. 21–27, Jan. 1967.

[15] C. R. Jack et al., "The Alzheimer's disease neuroimaging initiative (ADNI): MRI methods," *J. Magn. Reson. Imaging*, vol. 27, no. 4, pp. 685–691, Apr. 2008.

[16] Breiman, L. "Random forests," *Machine learning*, vol. 45, no. 1, pp. 5-32, 2001.

[17] M. A. Hearst, S. T. Dumais, E. Osuna, J. Platt, and B. Scholkopf, "Support vector machines," *IEEE Intell. Syst. Their Appl.*, vol. 13, no. 4, pp. 18–28, Jul. 1998.

[18] Y. Freund and R. E. Schapire, "A Decision-Theoretic Generalization of On-Line Learning and an Application to Boosting," *Journal of Computer and System Sciences*, vol. 55, no. 1, pp. 119–139, Aug. 1997.

[19] Guyon, Isabelle et al. "Gene Selection for Cancer Classification using Support Vector Machines." *Machine Learning*, vol. 46, pp. 389-422, 2004.

[20] H. Ding, P.-M. Feng, W. Chen, and H. Lin, "Identification of bacteriophage virion proteins by the ANOVA feature selection and analysis," *Mol. BioSyst.*, vol. 10, no. 8, pp. 2229–2235, 2014.

[21] M. Flint Beal, Michael F. Mazurek, Vinh T. Tran, Geetinder Chattha, Edward D.Bird, Joseph B. Martin, Reduced numbers of somatostatin receptors in the cerebral cortex in Alzheimer's disease, *Science*, vol. 229 (4710), pp. 289–291, 1985.